\documentclass[a4paper]{spie}  %>>> use this instead for A4 paper

\usepackage[]{graphicx}

\title{Advances in the Development of Mid-Infrared Integrated Devices for Interferometric Arrays}

\author{Lucas~Labadie\supit{a}, 
Guillermo~Mart\'in\supit{b}, 
Air\'an~R\'odenas\supit{c}, 
Norman~C.~Anheier\supit{d}, 
Brahim~Arezki\supit{b}, 
Robert~R.~Thomson\supit{c}, 
Hong~A.~Qiao\supit{d}, 
Pierre~Kern\supit{b}, 
Ajoy~K.~Kar\supit{c}, 
Bruce~E.~Bernacki\supit{d}
\skiplinehalf
\supit{a} I.\,Physikalisches Institut, Universit\"at zu K\"oln, Z\"ulpicher Str. 77, 50937 K\"oln, Germany \\
\supit{b} UJF-Grenoble 1/CNRS-INSU, Institut de Planetologie et d'Astrophysique de Grenoble (IPAG), UMR 5274, Grenoble, France\\
\supit{c} Scottish Universities Physics Alliance (SUPA), Institute of Photonics and Quantum Sciences, Heriot-Watt University, Edinburgh, EH14 4AS, United Kingdom\\
\supit{d} Pacific Northwest National Laboratory, 902 Battelle Boulevard, PO Box 999, Richland, 99352 Washington, USA
}

\authorinfo{Further author information: (Send correspondence to L.Labadie)\\L.Labadie: E-mail: labadie@ph1.uni-koeln.de, Telephone:\,+49 221 470 3493
}
\begin{document} 
  \maketitle 

%%%%%%%%%%%%%%%%%%%%%%%%%%%%%%%%%%%%%%%%%%%%%%%%%%%%%%%%%%%%% 
\begin{abstract}
This article reports the advances on the development of mid-infrared integrated optics for stellar interferometry. The devices are fabricated by laser writing techniques on chalcogenide glasses. Laboratory characterizaton is reported and analyzed. \end{abstract}

%>>>> Include a list of keywords after the abstract 


%%%%%%%%%%%%%%%%%%%%%%%%%%%%%%%%%%%%%%%%%%%%%%%%%%%%%%%%%%%%%
\section{INTRODUCTION}

The next decades will see the arrival of new large facilities like the Extremely Large Telescopes (ELTs) that will durably impact the observing capabilities in infrared astronomy. 
The advent of such gigantic facilities has triggered new ideas in designing the future instruments, in contrast with the conventional bulk optics designs, in which element size and cost scale with telescope size. 
% in rupture with the conventional bulk optics designs with size and cost scaling up with the telescope size. 
Photonics sciences can feed such new ideas. So far, photonics has been exploited for astronomical applications in few specific cases, such as for fiber-fed multi-object spectrograph. But the integration of a whole set of functions is somehow more recent. The synergy between the two fields gave birth to the recently highlighted field of ``astrophotonics''\cite{Bland-Hawthorn2009}\,. \\
The clear advantage is on the miniaturization of many instruments thanks to the fact that many optical functions can be integrated on a small chip and executed by means of waveguide couplers and switches, fibered remapping arrays, intensity and amplitude control, Bragg gratings optical filters and integrated spectrographs, or even new functions can be invented (e.g. pupil remapping). Typical dimensions of photonics components offer a great level of spatial integration, with consequent advantages on the cost, the production or the cryogeny of the instruments.
% Besides spectroscopy - e.g. with AWGs - and imaging with monolithic telescopes - e.g. with pupil remapping - 
One field that has greatly benefited from photonics devices is interferometry. 
%In the nineties and in the years 2000, interferometry has introduced new concepts for beam combination based on fiber couplers and on integrated-optics beam combiner. 
A particular characteristic of the photonics approach in interferometry is to rely on the single-mode behavior of the photonics components, which in addition to the strong level of instrument integration enables the modal filtering of the incoming wavefronts otherwise corrugated by the atmospheric turbulence. This permits a good calibration of the interferometric visibilities, helped in this by the simultaneous acquisition of the photometric signals.\\
In the last decade, photonics applications for interferometry has mainly been translated into the use of fibers and fiber couplers  (e.g. FLUOR\cite{Foresto1998,Foresto2003}\,, AMBER\cite{Petrov2007}\,, CHARA/MIRC\cite{Pedretti2009}), as well as of integrated optics components (IONIC/IOTA\cite{Berger2001}\,, PIONEER/VLTI\cite{LeBou2011}), producing high quality science results and unprecedented images at very high angular resolution. Relying on this success, future instruments such as GRAVITY at the VLTI are being constructed on the basis of an fiber-fed integrated optics beam combiner.\\
While the impact of photonics components on the ultimate science goals has been now widely recognized for the near-infrared bands, a transverse approach can be considered for the mid-infrared bands. Indeed, photonics components have been so far developed and adapted for stellar interferometry in the 1.2 -- 2.4\,$\mu$m wavelength range. This is a spectral domain where the silica-based technology -- mainly backed by the telecom industry -- is well mastered to produce reliable single-mode waveguides. In a recent review paper\cite{Labadie2009}\,, we underlined how the 2.4\,$\mu$m limit represent a real challenge for mid-infrared photonics: while the technical feasibility of such components has been sporadically proved via a variety of solutions, the lack of a stable and ``universal'' platform able to produce mid-infrared\,/\,single-mode\,/\,low loss\,/\,complex integrated optical circuits waveguides has so far limited the large expansion of the field of photonics in the 3--30\,$\mu$m.\\
However, mid-IR photonics would be of great interest for a high number of science cases in infrared astronomy, detailed elsewhere. 
% , and in particular in stellar interferometry where the strong level of integration would be a great asset when coming to cryogenic operations.
Historically, research on mid-infrared guided optics has also been actively supported by astronomers in view of future space-based interferometers. 
% devoted to exoplanetary science. 
This is one of the reasons we have pursued in these last years a technology program bringing together astronomers and photonicists to develop specific solutions for interferometry-oriented mid-IR integrated optics. The current status is presented in this paper.
%beyond its clear interest for ground-based interferometry.
%   Neighbor field : Gas sensing

\begin{figure}[b]
\centering
%\begin{minipage}[t]{\textwidth}
%\includegraphics[width=13.0cm]{composite3.ps}
%\end{minipage}
%\caption{{\it Left:} short-exposure image of the binary star GN Tau. The pattern is composed of different speckles with different brightness. {\it Center:} The seeing limited image of the source after a long exposure. The various speckles add randomly, resulting in an diameter independent FWHM of the PSF. {\it Right:} The shift-and-add plus frame selection image.}\label{image}
\end{figure}

\section{Accessing the mid-infrared wavelength range}% : a quick look in the past}

Fabricating waveguides operating beyond 2.4\,$\mu$m and implementing them on an astronomical facility has been tried -- sometimes with relative success -- for many years. On the astrophysics side, the TISIS experiment on IOTA was the only interferometric test showing the successful obtention of stellar fringes at 3.5\,$\mu$m using a X-shaped fluoride glass single-mode fiber coupler\cite{Mennesson1999}\,. While limited to the measurement of stellar diameters, it exhibited a remarkable instrumental stability. On the photonics side, the work done by different groups -- in particular outside the field of astronomy -- is much. We aim at summarizing hereafter some of the different research directions targeting the development of mid-IR guided optics. However, the field is so vast and diversified in its applications that we do not claim any exhaustivity. We rather attempt to put this research into the wider context of a possible and asserted interferometric interest.

\subsection{The fiber approach}

Fibers have been using in astronomy since many years for numerous applications such as multi-object spectroscopy, integral-field spectroscopy, high-precision radial velocity and interferometry. We only briefly discuss here the last case and refer to Heacox\,\&\,Connes\,(1992)\cite{Heacox1992} for a detailed review on the use of fibers in astronomy. As theoretically described by different authors\cite{Froehly1981,Tatulli2004}\,, {\it single-mode} fibers can improve the accuracy of the interferometric visibilities (down to 1\,\%) by mean of modal filtering, in particular when coupled to AO and in the low-light regime. This approach was successfully implemented on the IOTA interferometer\cite{Foresto1997} and on the succeeding near-IR facilities. In order to improve on the overall throughput, subsequent research programs have been funded to developed fluoride fibers optimized for the 2--2.4\,$\mu$m range.\\
At mid-infrared wavelengths, single-mode fibers have acquired high interest as wavefront modal filters\cite{Wallner2002} in the context of high-contrast nulling interferometry missions like DARWIN or TPF-I that would operate in the 4--20\,$\mu$m range. The requirement of high throughput coupled to the need of single-mode operation over a wide spectral range has raised interest among different groups who have proposed two type of technologies, namely silver-halide fibers\cite{Wallner2004,Ksendzov2008} and chalcogenide fibers\cite{Houizot2007,Aggarwal2002} based on research glasses to extend the wavelength coverage. Additional options are  hollow core fibers (Rutgers university/OKSI, Inc.) and photonic bandgap fibers. One should keep in mind that for such a wide spectral range these fibers need to be optimized for well-defined sub-bands\cite{Cheng2009}. Single-mode infrared fibers and hollow-core fibers\cite{Kriesel2011} are also available commercially, although they do not always offer a large spectral bandwidth of operation\cite{irphotonics}.
% or a single-mode behavior\cite{polymicro}.

\subsection{The integrated-systems approach}

As mentioned earlier, integrated optics is a branch of photonics offering a high level of compactness for many passive functions vital to astronomy and interferometry. Active devices such as LiNbO$_{\rm 3}$ modulators\cite{Heidmann2011,Caballero2012} are presently used as well, especially in the field of instrument calibration and control\cite{Follert2010}\,. Extending the concept of integrated photonics from the near to the mid-infrared (i.e. beyond 2.5\,$\mu$m and up to $~$20\,$\mu$m) is studied by different groups, including ourselves. Some directions are:
\begin{itemize}
	\item Silicon-on-insulator (SOI) technology. This is a technology platform which has been used for near-infrared photonic devices for more than twenty years, relying on low-loss properties of Silicon and SiO$_{\rm 2}$ in the near-infrared range. At wavelengths longer than 2.5\,$\mu$m, Silicon maintains a high transparency up to $\sim$7\,$\mu$m while SiO$_{\rm 2}$ losses increase rapidly beyond 3.6\,$\mu$m. If clearly not suited to the full mid-IR band, the SOI platform can remain a valid and technologically stable option for the astronomical L-band ($\sim$3--3.5\,$\mu$m). This has recently been demonstrated in the lab at 3.39\,$\mu$m with measured losses as low as 0.7\,dB/cm\cite{Mashanovich2011} on multimode waveguides. However, the drawback of the SOI approach for stellar interferometry is the physical cross-section of the waveguides required to be single-mode. Indeed, because of the strong index difference $n$(air-silicon-SiO$_{\rm 2}$), the fundamental mode alone would be supported for a waveguide cross-section of 0.6$\times$0.6\,$\mu$m, with obvious implications on the coupling efficiency with traditional optics. Increasing the operation range of the Silicon approach beyond 3.5\,$\mu$m has also been tested by implementing porous-Si as a substrate.
	\item An alternative recently advanced in order to increase the transmission range of the waveguides  is Silicon-on-Sapphire. With a good transparency up to 6--7\,$\mu$m and low index, Sapphire is interesting to be used as the low-index substrate\footnote{In the SOI technology, the waveguide is structured as Si(high-index core)/SiO$_{\rm 2}$(low-index cladding)/Si(high-index substrate). The SiO$_{\rm 2}$ thin layer confines the guided mode, but because of its few microns thickness it can also induce mode leaking in the Si substrate.}, which also reduces the risk of substrate leaking. This technology has been experimentally tested at 4.5\,$\mu$m and 5.18\,$\mu$m\cite{BaehrJones2010,Li2011}\,, exhibiting propagation losses of 4.3\,dB/cm and $<$2\,dB/cm, respectively, for single-mode structures.
	\item More exotic solutions based on hollow core metallic waveguides were studied by ourselves, with the objective of exploiting the propagation in the air in order to enlarge the operational spectral range. However, the experimental results revealed significantly high-losses\cite{Labadie2006}\,, which makes this solution more adapted for ultra-short horned channel waveguides to be used as spatial filters rather than for complex integrated optics functions, or to be used at far-IR wavelengths.	
	\item A fourth and promising solution is covered by the chalcogenide glass waveguides technology. Chalcogenide glass is a popular infrared amorphous glass transparent up to $\sim$12\,$\mu$m, which transparency can be ``tuned'' via the production of telluride compositions. Chalcogenide single-mode rib-waveguides have been produced using etching techniques\cite{Vigreux2007,Vigreux2011}\,. However, a very promising approach to those glasses resides in the {\it laser writing} technique, a method that exploits the permanent photo-darkening effect to produce more complex integrated functions. In the next sections, we report our recent results using this approach.
\end{itemize}

\section{Waveguide fabrication}

\subsection{Principle of laser writing}

\begin{figure*}[b]
\centering
%\begin{minipage}[t]{\textwidth}
\includegraphics[width=5.0cm]{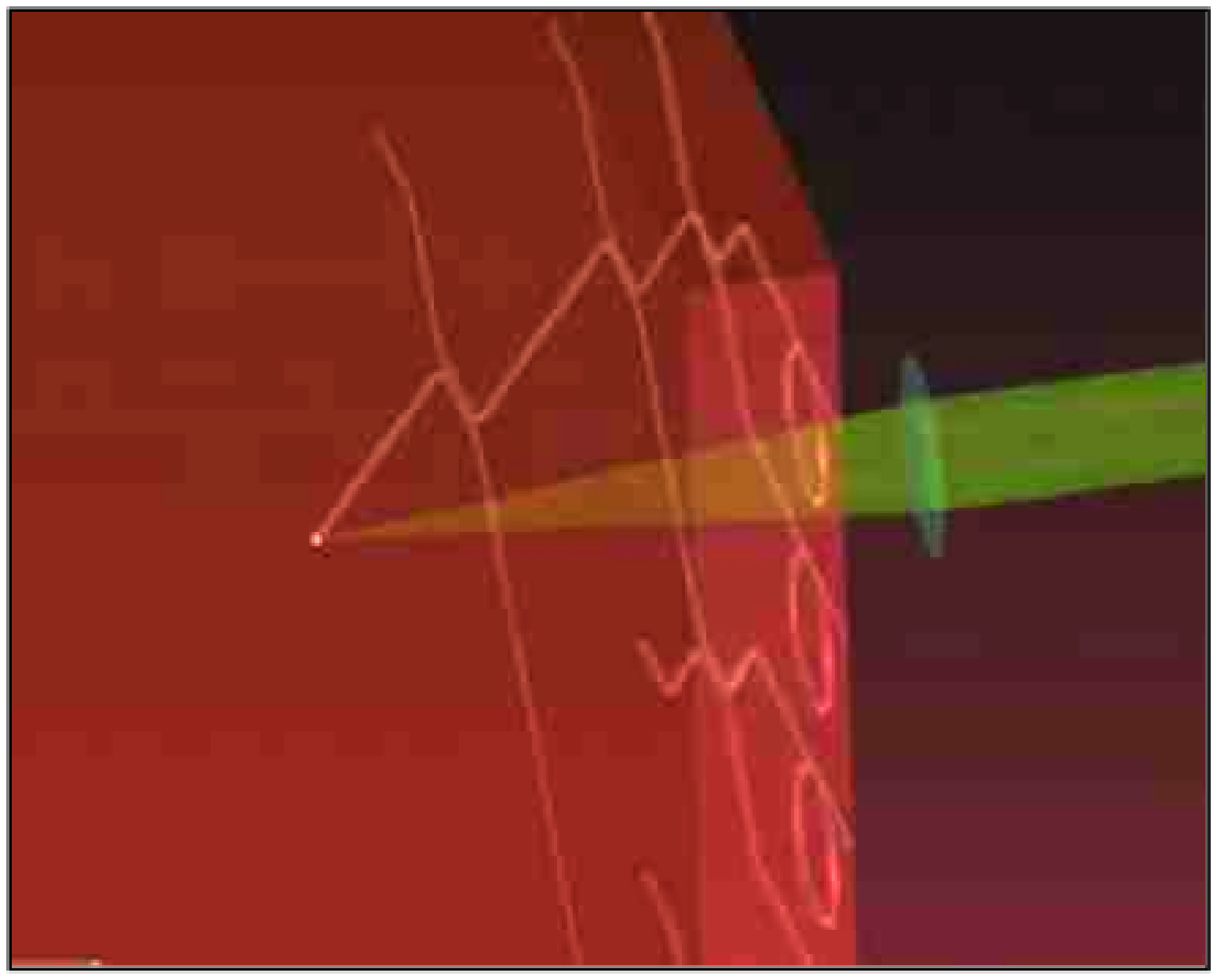}
\hspace{1.4cm}
\raisebox{0.0cm}{\includegraphics[width=4.5cm]{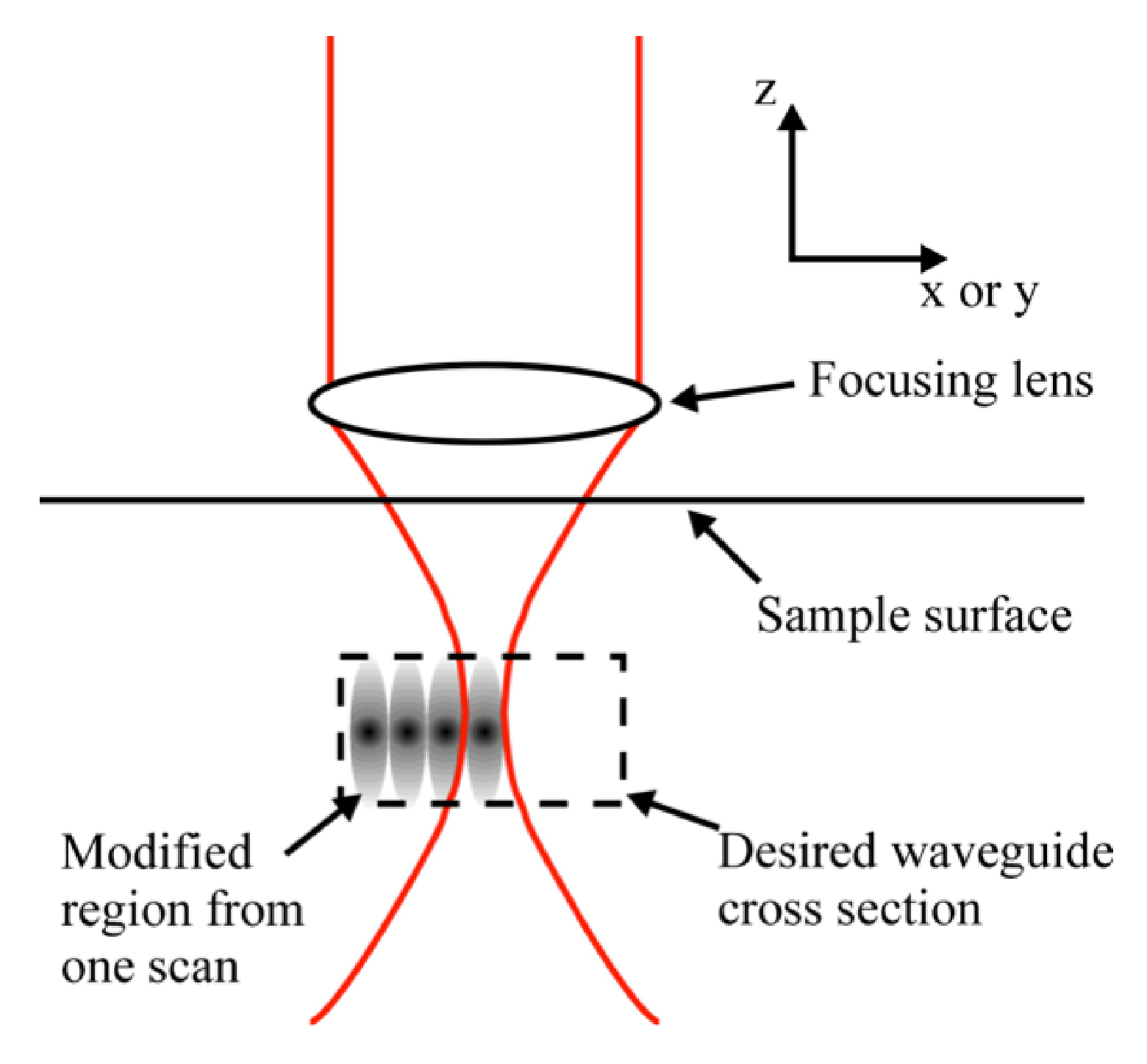}}
%\vspace{0.1cm}
%\raisebox{0.0cm}{\includegraphics[width=6.5cm]{images/structure.ps}}
%\hspace{1cm}
%\raisebox{0.0cm}{\includegraphics[width=6.5cm]{images/structure.ps}}
%\end{minipage}
\caption{Laser writing technique. {\it Left: } Cartoon illustrating the two-dimensional writing of waveguides using a CW HeNe laser on a As$_{\rm 2}$Se$_{\rm 3}$ core layer. {\it Right: } Principle of the three-dimensional ultrafast laser inscription technique using a pulsed laser (from Thomson et al. (2012)\cite{Thomson2012}).}\label{Fig1}
\end{figure*}

This technique was already investigated by Efimov et al. (2001)\cite{Efimov2001} to write waveguides in chalcogenide glasses at 850\,nm. The principle consists in focusing a high-power external radiation source like a laser onto a given location on the substrate surface in order to increase the refraction index of the glass, creating therefore a spatially localized index increase able to confine a wave. Laser writing can be obtained using a CW He-Ne laser or a ultrashort pulsed laser (e.g. at $\lambda$=850\,nm or 1047\,nm) as the writing probe. The approach is slightly different in the two cases and results essentially in accessing -- or not -- the third vertical direction for writing waveguides. This is illustrated in Fig.~\ref{Fig1} and explained hereafter. The left-hand cartoon shows how the CW HeNe laser can pattern a structure of waveguides on the surface of a chalcogenide glass thin layer. The right-hand cartoon illustrates how a pulsed laser can be focused and displaced over the {\it three} directions $X$, $Y$, $Z$ within a chalcogenide substrate, hence producing bent waveguides to be potentially gathered into a ``waveguide bundle''. A specific {\it modus operandi} called multi-scan technique permits us to taylor independently both the waveguides size and index. The multi-scan technique produces square-section or rectangular-section waveguides.

\begin{figure*}
\centering
%\begin{minipage}[t]{\textwidth}
\includegraphics[width=15cm]{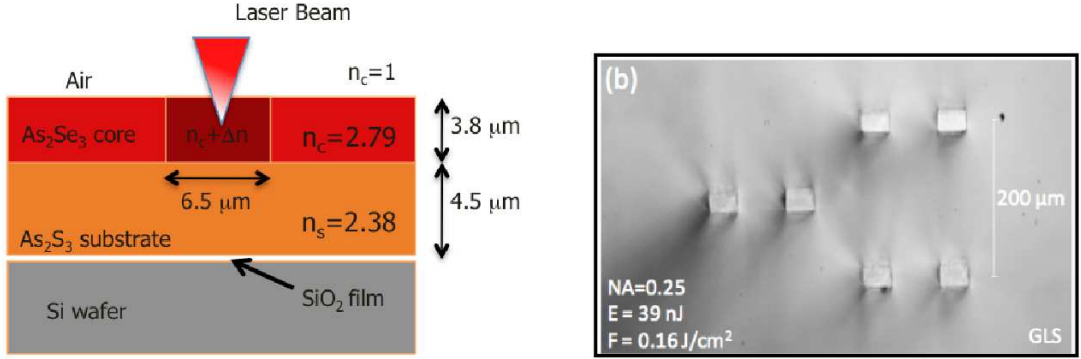}
%\end{minipage}
\caption{{\it Left: } Cross-section geometry of the two-dimensional laser written Y-junction delivered by the PNNL, with characteristic sizes of the waveguide. {\it Right: } View of the waveguides fabricated by ultrafast laser inscription technique. The spatial distribution of the waveguides shows how the three spatial directions are exploited.}\label{Fig2}
\end{figure*}

\subsection{Two-telescope Y-junction beam combiner}

A Y-junction beam combiner has been produced by the PNNL\footnote{Pacific Northwest National Laboratory} using the CW HeNe laser writing technique. 
The design foresees an asymmetric-strip embedded waveguide structure. In the vertical direction, the high-index waveguide core is bounded by air (the {\it upper cladding}) and As$_{\rm 2}$S$_{\rm 3}$ (the {\it lower cladding}). The core is made of high-index photomodified As$_{\rm 2}$Se$_{\rm 3}$. Because of the vertical index asymmetry for the superstrate--core--substrate arrangement, the vertical distribution of the fundamental mode is slightly asymmetric, which can result in coupling degradation when overlapped with the radially symmetric injection beam. In the horizontal direction, the waveguide core is symmetrically bounded by lower-index unmodified As$_{\rm 2}$S$_{\rm 3}$ regions. The waveguide parameters are the thickness $d$=3.8\,$\mu$m and the width $w$=6.5\,$\mu$m (cf. Fig.~\ref{Fig2}). Considering the refractive index values of As$_{\rm 2}$Se$_{\rm 3}$ and As$_{\rm 2}$S$_{\rm 3}$ available in the literature, the structure is single-mode by design beyond $\lambda$=8.7\,$\mu$m. The fundamental mode has an elliptical shape with minor and major axes of $\sim$9\,$\mu$m and $\sim$18\,$\mu$m, respectively.
For the manufacturing phase, a first step is the consecutive thermal deposition of a 3.8\,$\mu$m thick film of As$_{\rm 2}$Se$_{\rm 3}$ on a 4.5\,$\mu$m thick film of As$_{\rm 2}$S$_{\rm 3}$, according to the deposition parameters of H\^o et al.(2006)\cite{Ho2006}\,. At 633\,nm, the photons energy is larger than the bandgap of As$_{\rm 2}$Se$_{\rm 3}$, enabling the absorption of laser  photons by the core layer. Non-linear phenomena under high irradiation results in a localized modification of the glass structure with a consequent modification of the refractive index. The high-index core layer can hence absorb those photons and be photomodified. At the contrary, this is not the case for the As$_{\rm 2}$S$_{\rm 3}$ substrate, which bandgap is higher than the photon energy. 
The refractive indices for the substrate As$_{\rm 2}$S$_{\rm 3}$ and the core layer As$_{\rm 2}$Se$_{\rm 3}$ are, respectively, $n_{\rm clad}$=\,2.38 and $n_{\rm core}$=\,2.78 at 10\,$\mu$m, while the laser-writing process induces a local index increase of 0.04--0.05\cite{Ho2006}\,. The manufactured Y-junction is $\sim$30\,mm long with a separation of 7\,mm between the two arms.
% The cartoon in Fig.~\ref{Fig1}-left shows the case where a CW He-Ne laser is focused on the surface of a few micrometers thick As$_{\rm 2}$Se$_{\rm 3}$ chalcogenide layer. 
% The main -- as well as major -- difference when using one or the other type of source is on

\subsection{Three-telescope and three-dimensional beam combiner}

\begin{figure*}[t]
\centering
%\begin{minipage}[t]{\textwidth}
\includegraphics[width=13cm]{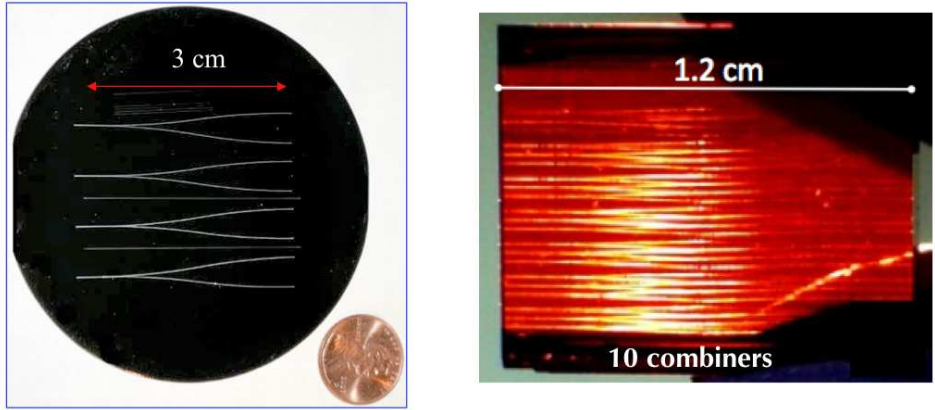}
%\end{minipage}
\caption{{\it Left: } two-dimensionally written Y-junctions in Arsenic-based glass. {\it Right: } three-dimensionally written 3T beam combiners in Sulphide glass.}\label{Fig3}
\end{figure*}

A second approach we investigated is the three-dimensional laser-writing of waveguides. Here, a pulsed laser is used as the writing tool and translated three-dimensionally directly into a chalcogenide glass bulk. Hence no preliminary high-index core layer deposition is required. Pulsed laser can deliver extremely high peak intensities over a few hundred of femtoseconds. Although the photon energy is sub-bandgap, the peak power reached at the laser focus is high enough to induce multi-photon absorption, resulting in a local modification of the material and of its index as this is shown in Fig.~\ref{Fig1}. The multi-scan technique allows one to define independently the rectangular cross-section of the waveguide.\\
We have used research and commercial free-Arsenic chalcogenide sulphide glasses. The transmission upper limit is about $\sim$10--11\,$\mu$m for both type of bulk glasses while the absorption is higher compared to Arsenic-based compounds such as As$_{\rm 2}$S$_{\rm 3}$ (on the order of 2\,dB/cm). Different waveguiding structures were laser-written using the pulsed laser with the parameters of Rodenas et al. (2012)\cite{Rodenas2012}\,. We achieved $\Delta n$ ranging from $\sim$0.008 to $\sim$0.012, respectively for the single-mode and multimode structures at 10\,$\mu$m, i.e. lower than in the case of CW laser-writing. The waveguides have a cross-sectional diameter of circa 35\,$\mu$m and a mode field diameter (MFD) varying from 53\,$\mu$m to 48\,$\mu$m. An image of the fabricated waveguides is shown in Fig.~\ref{Fig2}. We have then produced a set of {\it three-telescope} beam combiners based on Y-junctions as shown in the right panel of Fig.~\ref{Fig3}. However, these devices were characterized later only as independent two-telescope Y-junctions due to the current design of the testbeds.\\
\\
Two photographs of the fabricated devices are shown in Fig.~\ref{Fig3}. These devices were designed to be single-mode at 10.6\,$\mu$m, with a bi-mode/single-mode cutoff between 8 and 9\,$\mu$m.

\section{Characterization and results}

\subsection{Goal of the characterization phase and testbed description}

\begin{figure*}[t]
\centering
%\begin{minipage}[t]{\textwidth}
\includegraphics[width=13cm]{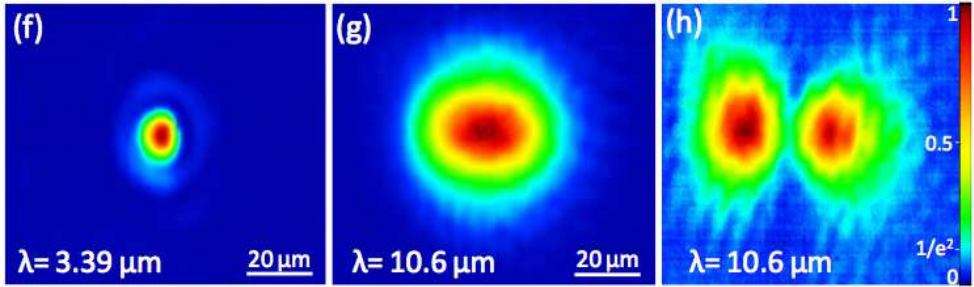}
%\end{minipage}
\caption{Near-field imaging characterization {\it Left: } at 3.39\,$\mu$m. {\it Middle and Right: } Effect on the near-field of single-mode versus multimode behavior. In this last case, the first higher order mode can be excited depending on input coupling conditions.}\label{Fig3b}
\end{figure*}

A first characterization step is to analyze the modal behavior of the newly fabricated devices, and possibly to identify the different spectral intervals where the component can be single-mode, bi-mode etc... There are typically two ways of proceeding. \\
The first one is to couple light into the input of the test waveguide using a fast lens or a single-mode fiber if available, and to image the output on a camera in such a way that the spatial profile of the modes is resolved. When apply small lateral displacements to the input beam, the different modes of a multimode waveguide can be excited and their characteristic shape clearly observed. For a single-mode waveguide, only the fundamental mode is resolved with its overall intensity varying as a function of the input coupling conditions. This method generally uses a monochromatic source and provide us information on the waveguide modal behavior at a given wavelength. An example is given with Fig.~\ref{Fig3b}, where the single-mode versus multimode behavior for some three-dimensional ULI samples is assessed at 3 and 10\,$\mu$m.\\
The second method is based on the analysis of the waveguide transmission spectrum normalized by the transmission spectrum of the testbench. We know from the theory that for a given waveguide, the number of supported modes decreases with increasing wavelength. Hence, the high-order propagation modes become gradually cut-off as the radiation wavelength increase, until the fundamental mode alone is still propagated. This loss in the number of supported modes can be observed in the form of sharp drops in the normalized transmission spectrum which correspond to partial loss of transmitted power. 
%-----------------------
%-----------------------
Due to its polychromatic nature, the {\it transmitted power method}\,\cite{Lang1994} help us to identify the different spectral ranges of a corresponding spectral regime.\\
\\
A second important characterization step is the analysis of the interferometric performances of the fabricated device. Depending on the beam combiner feature (two- or three-telescope combiner) we aim at quantifying important parameters such as
\begin{itemize}
\item the monochromatic and polychromatic interferometric contrast.
\item the instrumental phase and the instrumental closure phase
\item the photometric unbalance
\item the total throughput, the propagation losses
\item the chromatic dispersion for wide-band operation
\item input and output coupling losses
\item differential polarization effects between the different channels
\end{itemize}
Other aspects may concern the temporal stability of the previous parameters, the response of the device to cryogenic temperatures, the impact of aging effects, or the mechanical robustness. Although it is not always possible to characterize all these parameters, this is likely the ``minimum package'' to be assessed in preparation to on-sky operation. \\
\\
For the purpose of this work, we have developed a spectroscopic testbed for the modal characterization and a Michelson-type testbed to be used for the interferometric characterization and/or for monochromatic near-field imaging. 
The first bench is a Fourier Transform Spectrometer (FTS) designed to couple light in and out of the sample to be characterized. It covers the spectral range 2--14\,$\mu$m. The interferometric bench can operate in monochromatic light ($\lambda$=10.6\,$\mu$m) and with white-light sources as well, and can currently host only two-telescope beam combiners. More details can be found in Labadie et al. (2011)\cite{Labadie2011}\,. 
% -------------------------
\begin{figure*}
\centering
%\begin{minipage}[t]{\textwidth}
%\includegraphics[width=5.0cm]{images/2D-writing.ps}
%\hspace{1.4cm}
\raisebox{0.0cm}{\includegraphics[width=8.2cm]{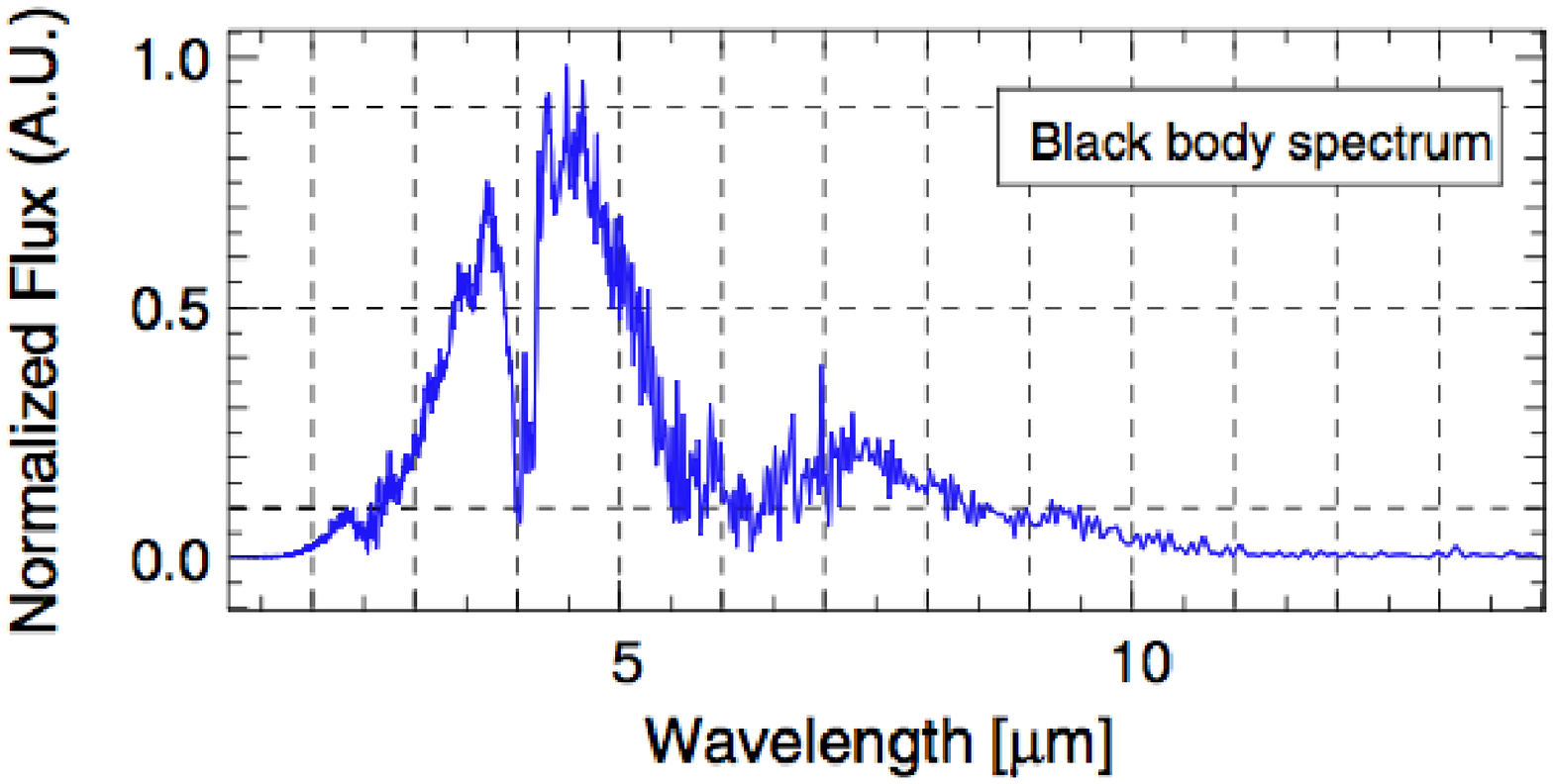}}
\hspace{0.5cm}
\raisebox{0.0cm}{\includegraphics[width=8.2cm]{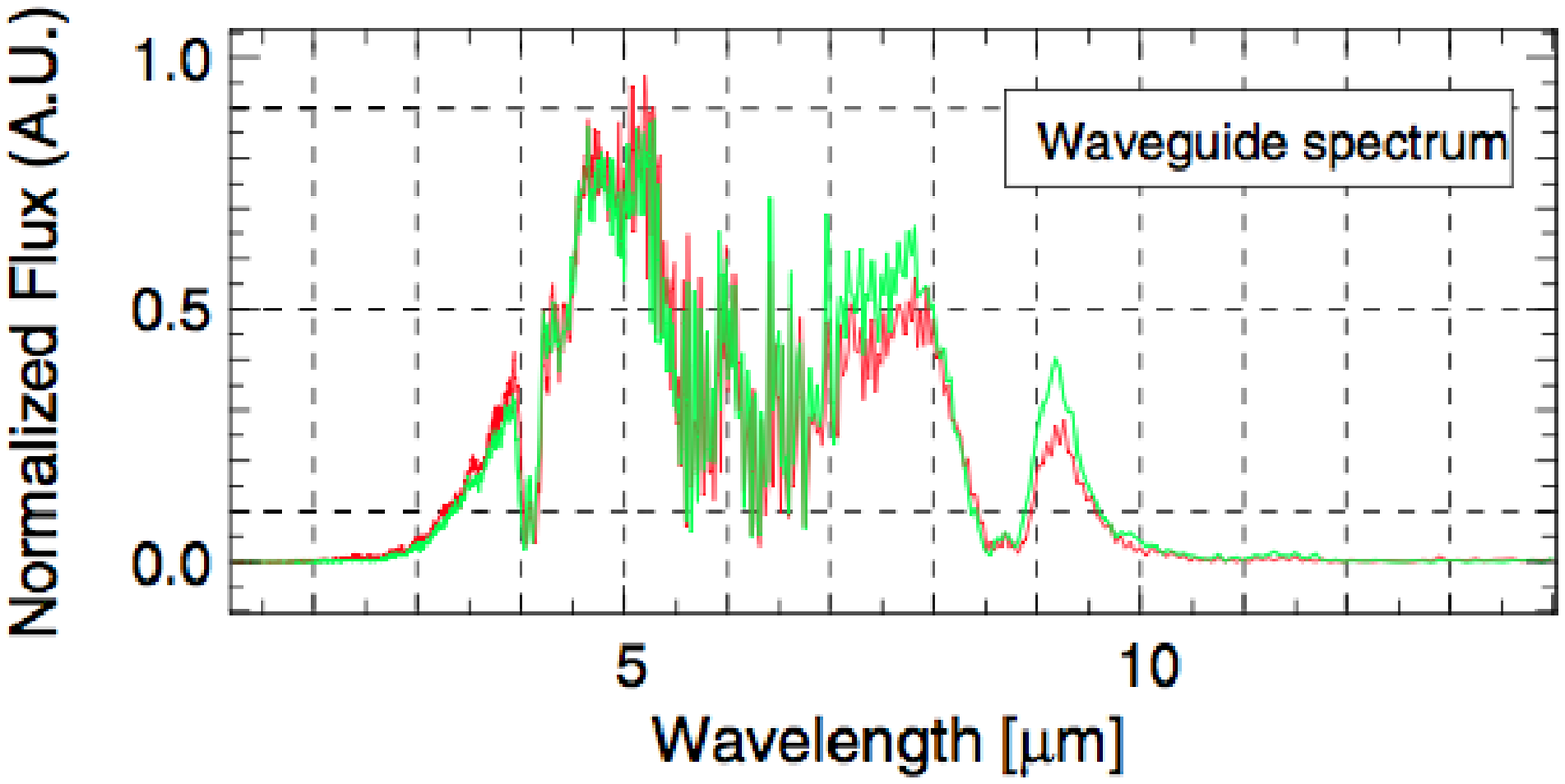}}
%\hspace{1cm}
%\raisebox{0.0cm}{\includegraphics[width=6.5cm]{images/structure.ps}}
%\end{minipage}
\caption{Waveguide and calibration transmission spectra showing the possible bi-mode/single-mode cutoff at 8.5\,$\mu$m. From Labadie et al. (2011)\cite{Labadie2011}\,.}\label{Fig4}
\end{figure*}
% -------------------------

\subsection{Modal characterization}

In addition to the near-field imging technique, we have investigated with the FTS testbed the modal behavior of the chalcogenide channel waveguides that were fabricated under the same conditions of the Y-junction delivered by the PNNL. The objective was to identify the bi-mode/single-mode cutoff theoretically predicted in the 8--12\,$\mu$m range. We recorded multiple transmission spectra of the component between 2 and 14\,$\mu$m together with several calibration spectra of the whole system without waveguide in order to eliminate the spectral response of the bench. A grid polarizer placed in the optical path permitted us to investigate both TE and TM polarization. The comparison of the two spectra is shown in Fig.~\ref{Fig4}. The spectra are affected by CO$_{\rm 2}$ absorption at 4.2\,$\mu$m and by water vapor absorption between 5 and 7\,$\mu$m, where any quantitative measurement becomes irrelevant. The FTS interferograms are scanned over 2\,mm, resulting in 5\,cm$^{-1}$ spectral resolution. The blue curve in the left panel is the blackbody calibration raw emission spectrum acquired within the sensitivity range of the MCT detector, and which includes absorption features due to the lab atmosphere and to the testbench components. The red and green curves are, under identical experimental conditions, two different instances of the channel waveguide spectrum after coupling light in it. Both spectra have been normalized to their peak intensities since the two curves cannot be directly compared: only the {\it relative} variations are sought here.\\
The clear intensity drop observed at 8.5\,$\mu$m is interpreted as the bimode/single-mode cut-off wavelength of the channel waveguide, which is also in good agreement with the theoretical value.
% -------------------------
\begin{figure*}[h]
\centering
%\begin{minipage}[t]{\textwidth}
%\includegraphics[width=5.0cm]{images/2D-writing.ps}
%\hspace{1.4cm}
\includegraphics[width=17cm]{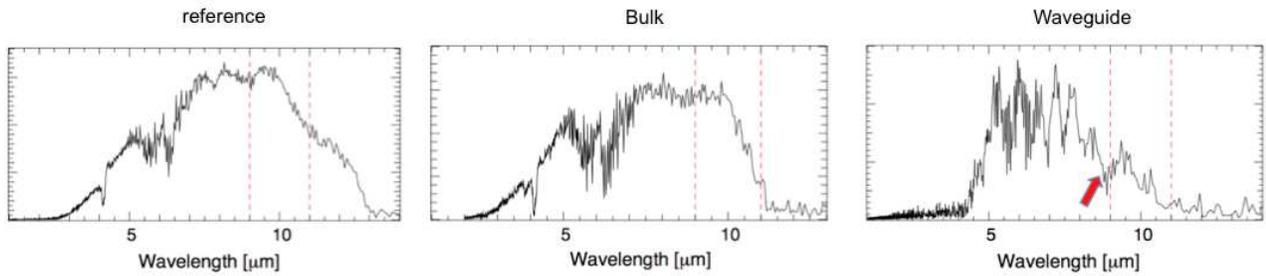}
\caption{Waveguide and calibration transmission spectra for the three-dimensional ULI written samples. The red arrow indicates the possible cutoff wavelength.}\label{Fig5}
\end{figure*}\newline
% -------------------------
We conducted an analog study on the samples fabricated by ultrafast laser inscription. 
In this case, we compared qualitatively the trend of the transmission spectrum of the reference (i.e. bench without any component), of the bulk, and of the waveguide. This allows a normalization by the spectral response of the bench and/or of the bulk. The intensity on the y-axis is therefore in arbitrary units and not displayed here. The spectra presented in Fig.~\ref{Fig5} covers the 2-14\,$\mu$m wavelength range and exhibit clear differences. In particular, the bulk spectrum shows a transparency cutoff around 11\,$\mu$m of the GCIS glass composition (75GeS$_{\rm 2}$-15Ga$_{\rm 2}$S$_{\rm 3}$-4CsI-2Sb$_{\rm 2}$S$_{\rm 3}$-4SnS) as reported in Rodenas et al. (2012)\cite{Rodenas2012}\,. Concerning the transmission spectrum of the waveguide, the trend is affected by water vapor absorption between 5 and 7\,$\mu$m and exhibit several peaks afterwards up to 9\,$\mu$m which are more difficult to interpret. However, we observe at 9\,$\mu$m a sharp cut-off followed by a decrease in intensity which appears very characteristic of a the bi-mode/single-mode cutoff of the waveguide (indicated by a red arrow).\\
Complemented by interferometric tests, the spectroscopic analysis reinforce our statement of single-mode waveguides successfully fabricated for the 10\,$\mu$m range. 

\subsection{Interferometric characterization}

\subsubsection{Monochromatic tests}

\begin{figure*}[h]
\centering
\includegraphics[width=17cm]{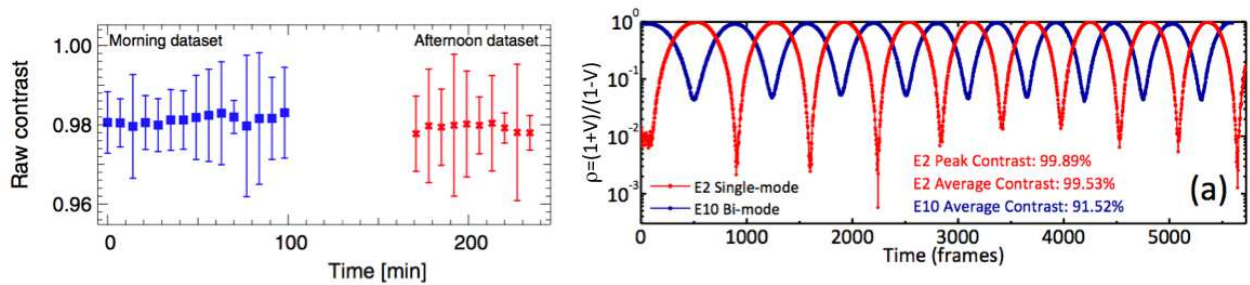}
\caption{Monochromatic interferometric characterization of the two types of devices.}\label{Fig7}
\end{figure*}

The first set of interferometric measurements aims at evaluating the achievable monochromatic contrast with the newly fabricated 2-D and 3-D devices. For this test, we use an adapted Michelson interferometer scheme where the two phase-shifted beams are injected separately into each of the inputs of a Y-junction function. By scanning the delay-line over typically ten fringes, we record the nulled and bright output from which the contrast is measured, after correction of the photometric unbalance between the two channels. The interferometric contrast and corresponding uncertainties are computed as described elsewhere\cite{Labadie2011}\,.\\
We measured the contrast for three different devices: the {\it single-mode} Y-junction fabricated via 2-D CW laser writing by the PNNL; a {\it single-mode} Y-junction fabricated via 3-D ultrafast laser inscription; a {\it multi-mode} Y-junction fabricated following the same process. The result are presented in Fig.~\ref{Fig7}. In general, the monochromatic nature of the interferences ensures a relatively high contrast (90\% or higher), whatever the quality of the wavefront filtering is\footnote{The polychromatic nature of the interference results generally in a degradation of the contrast.}. It is therefore required to seek for differences in contrast at the 1\% level or better when operating with laser light.
%-------------------------
\newline
%-------------------------
\noindent In the left panel of Fig.~\ref{Fig7}, the stability of the interferometric contrast obtained with the PNNL sample is plotted against time over a duration of several hours. 
The contrast appears to remain very stable against the fluctuations of the laboratory environment. The mean contrast is measured at $C$=0.9807$\pm$0.0003 over 300\,min. The 1-$\sigma$ dispersion of the contrast is $\sigma$=0.001 (0.1\%), and the propagated error from the individual measurements is $\epsilon$=0.009 (0.9\%).\\
The right panel of Fig.~\ref{Fig7} shows a direct comparison of a multimode and a single-mode component obtained using 3-D ultrafast laser inscription. The single-mode contrast is once again very high with an average value of 99.5\% and a peak value of 99.89\% (or $\rho$=5$\times$10$^{-4}$ in terms of rejection ratio). The contrast measurement in the case of the multimode structure leads to an average contrast of ``only'' 91.5\%, or almost one order of magnitude worse in terms of rejection ratio.\\
These results clearly shows the impact of the single-mode integrated optics device on the final quality and accuracy of the instrumental contrast.

\subsubsection{Polychromatic tests}

\begin{figure*}[h]
\centering
\includegraphics[width=17cm]{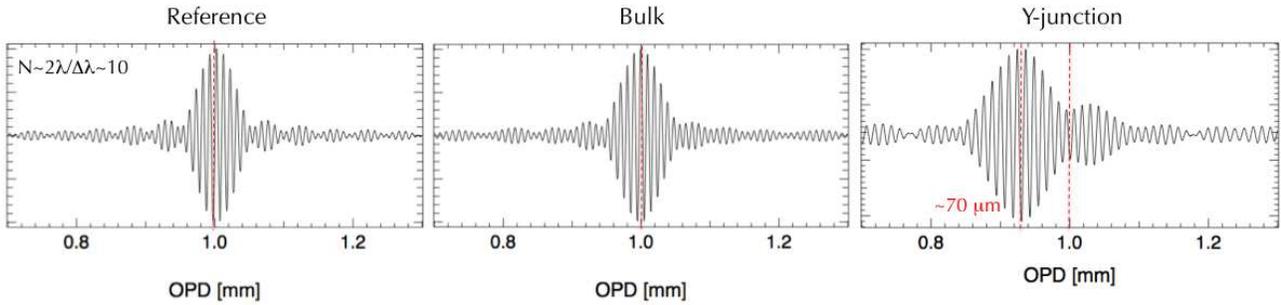}
\caption{White-light interferograms in the 9--11\,$\mu$m range obtained from numerical filtering of the FTS spectra. The comparison is made between the reference (bench), the bulk, and the beam-combiners.}\label{Fig8}
\end{figure*}

Interferometric polychromatic tests are also required in order to determine the achievable white-light contrast (i.e. close to on-sky operation) and to evidence possible chromatic effects related to dispersion. In this work, we did not use a hardware bandpass filter centered on 10\,$\mu$m, which would have permitted to directly record the interferometric and photometric signals, correct for the flux unbalance and derive the absolute contrast. Here we derive the shape of the interferogram out of the various recorded spectra shown in Fig.~\ref{Fig7} as follows. The spectra are numerically filtered with a top-hat filtered centered on 10\,$\mu$m and with a bandwidth $\Delta\lambda$=2\,$\mu$m. The filtered spectral range is represented by two dashed vertical lines on the different plots of Fig.~\ref{Fig8}. The Fourier Transform of the filtered spectra gives then the white-light interferogram for a 20\% bandwidth centered at 10\,$\mu$m. Note that only the spread of the interferogram is meaningful in this case in order to interpret possible dispersion effect. The absolute interferometric contrast (or visibility) cannot be retrieved since no photometric signal is available in the adopted bandwidth. Hence this important piece of information needs to be measured elsewhere.\\
% ---------------------
% ---------------------
% ---------------------
% ---------------------
\\
\noindent The left- and middle-panel of Fig.~\ref{Fig8} show the white-light interferograms obtained as a reference (i.e. bench without any sample), as well as through the GCIS bulk. 
The fringe packets appear quite symmetric and the number of fringes is about ten, in agreement with the theoretical number of fringes expected for a white-light interferogram where no dispersion occurs ($N$$\sim$2$\lambda_{\rm c}$/$\Delta\lambda$, where $N$ is the number of fringes and $\lambda_{\rm c}$ the central wavelength). In the case of the Y-junction, the right-panel of Fig.~\ref{Fig8} shows a significantly different behavior of the fringe packet, which has a higher spread in OPD. Two systems of fringes can be observed, which is a clear indication of dispersion. The main packet is shifted by circa 70$\mu$m with respect to the reference position. Typically, three possible origins are considered for such a level of dispersion: i) differential length of between two arms of the Y-junction and consequent chromatic dispersion. ii) modal dispersion if the device is multimode or if the two arms have slightly different indices of refraction (i.e. the propagation constants are not strictly equal). iii) strong induced birefringence resulting in different behaviors for TE and TM polarization. Currently, the origin of the observed dispersion effect is still under investigation. However, some first reasoning is not in favor of the i) hypothesis (differential length between the two arms) since the laser location in the ultrafast laser inscription technique can be controlled down to $\pm$20\,nm, although this hypothesis cannot be discarded so far. In the case of the ii) hypothesis, typical differences between the indices of refraction of the two arms are on the order of $\Delta$n$\sim$10$^{\rm -4}$. A simple theoretical estimate of the resulting dispersion (i.e. $\Delta$n.$L$, where $L$ is the length of the arms) is not in agreement with a 70\,$\mu$m shift of the central fringe packet. Concerning the last hypothesis (i.e. birefringence) this has not been investigated yet, but if any, this effect could in a large part be responsible for such an effect. It is likely that it is, at the end, a combination of the three effects that results in the dispersion of the white-light interferogram.\\
Note however that, if operating on an instrument, such devices would be coupled to a spectrometer, which by definition reduces the spectral bandwidth and diminishes the effect of the dispersion.

\subsection{Throughput and propagation losses}

As a complementary measurement, with provided some first estimates of the transmission characteristics of these new devices. Both the normalized spectrum technique and the Fabry-Perot etalon method were used independently. With the first technique an estimate of the coupling losses is needed in order to decouple them from propagation losses, which renders the final estimate more uncertain. The second method provides a more accurate measurement of the propagation losses.\\
For the 2-D laser written device fabricated by the PNNL, independent losses were measured elsewhere on similar waveguides at the level of 0.5\,dB/cm, which is a very low number in view of future use at the telescope. This is because laser-written waveguides do not introduce the type of scattering losses found in etched waveguides.\\
% ---------------------
% ---------------------
\begin{figure*}[h]
\centering
\includegraphics[width=17cm]{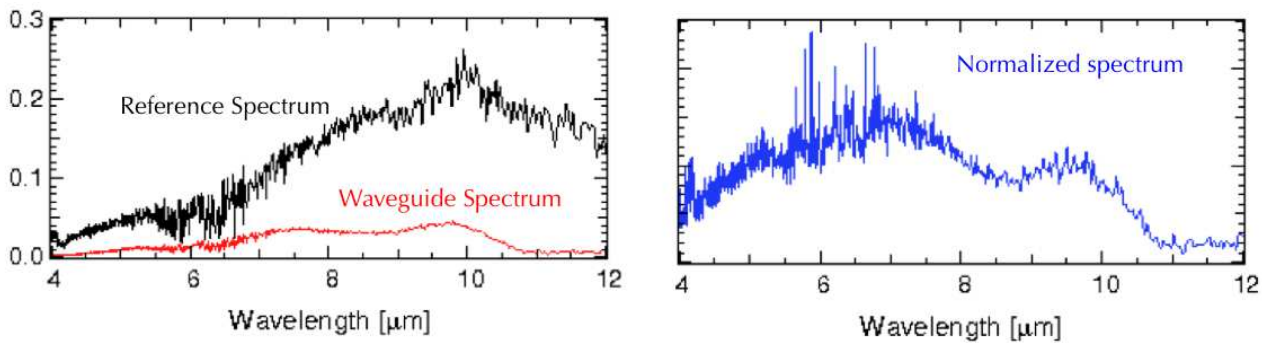}
\caption{Calibrated transmission spectrum for the estimate of the total throughput of the 3-D samples in the 7--10\,$\mu$m range.}\label{Fig9}
\end{figure*}
% ---------------------
% ---------------------
\\
The 3D ULI single-mode sample were first measured by spectroscopy as illustrated in Fig.~\ref{Fig9}. The normalized spectrum gives an indication of the total throughput in the spectral range 7--10\,$\mu$m. A 20\% throughput estimate is derived within the single-mode spectral range. This number accounts also for the Fresnel losses, which are at the level of $\sim$35\% in and out for uncoated facets. If the input and output coupling efficiency is made to vary between 65\% and 78\% the resulting propagation losses are found to vary between 0.7\,dB/cm and 2.2\,dB/cm, respectively. The lower value in this range is in agreement with what found by Ho et al. (2006)\cite{Ho2006} on Arsenic-based components. However, recent Fabry-Perot etalon measurements conducted on the ULI fabricated waveguides suggest that the propagation losses could be more on the order of 2--2.5\,dB/cm. Further investigation is ongoing in order to optimize the process.

\section{Conclusions}

We reported in this paper recent advances made by our group in the field of mid-infrared integrated devices for astronomical applications, and in particular for telescopes beam combination in stellar interferometry. We developed both Y-junctions and three-telescope beam combiners based on the laser inscription of infrared glasses like Chalcogenide glasses. Very encouraging results are reported in terms of beam combination, single-mode behavior, stability of high-level monochromatic instrumental visibilities, and first estimate of the propagation losses. Further efforts are foreseen to investigate the performances of these devices in white-light operation, and in particular to investigate the origin of the measured dispersion of the fringe packets for a 20\% bandwidth. On the short-term, we envision similar developments optimized for the L band, around 3.8\,$\mu$m.

%%%%%%%%%%%%%%%%%%%%%%%%%%%%%%%%%%%%%%%%%%%%%%%%%%%%%%%%%%%%%

%%%%%%%%%%%%%%%%%%%%%%%%%%%%%%%%%%%%%%%%%%%%%%%%%%%%%%%%%%%%%
\acknowledgments          
 
This work was supported by the U.S. Department of Energy, Office of Nonproliferation and Verification Research and Development (NA-22). Pacific Northwest National Laboratory is operated for the U.S. Department of Energy by Battelle Memorial Institute under Contract No. DE-AC05-76RLO1830.\\
A. R{\'o}denas acknowledges financial support from the Spanish Ministerio de Educación under the Programa de Movilidad de Recursos Humanos del Plan Nacional de I+D+I 2008/2011 for abroad postdoctoral researchers.

%%%%%%%%%%%%%%%%%%%%%%%%%%%%%%%%%%%%%%%%%%%%%%%%%%%%%%%%%%%%%
%%%%% References %%%%%

\bibliography{report}   %>>>> bibliography data in report.bib
\bibliographystyle{spiebib}   %>>>> makes bibtex use spiebib.bst

\end{document}